# A temporal proximity network dataset from a wedding cocktail hour with relationship category labels

Joshua Z. Stadlan[1,2,3,4 *], Richard B. Kahn[5], Michelle Birkett[1,2, 3]

[1] Center for Computational & Social Sciences in Health, Northwestern University, Chicago, IL, USA
[2] Northwestern Institute on Complex Systems, Northwestern University, Evanston, IL, USA
[3] Department of Medical Social Sciences, Feinberg School of Medicine, Northwestern University, Chicago, IL, USA
[4] Kellogg School of Management, Northwestern University, Evanston, IL, USA
[5] Independent researcher, Chicago, IL, USA
[*] Corresponding author. Email: joshua.stadlan@kellogg.northwestern.edu; ORCID: 0000-0003-4485-6778

## Abstract

**Objectives:** We captured a fine-grained dataset of unstructured social interaction with socially meaningful group labels to fill a gap in the study of face-to-face interaction. Prior interaction data from conferences, classrooms, hospitals, and workplaces exhibit network signatures such as heterogeneous contact rates, clustering, and bursty dynamics. However, schedules, room assignments, and authority roles in these settings may obscure unstructured social group dynamics. Studies on group mixing often rely on demographic proxies like gender, or assigned categories like school classes, rather than relationship-based groups. We aim to understand if temporal network signatures of institutionally structured settings generalize to unstructured social interaction.

**Data description:** We present the first public temporal proximity network dataset of a privately hosted social event with contextual relationship-based group membership. At the outdoor cocktail hour of a wedding, 95 participants wore proximity sensor badges that detected other badge-wearers within approximately 1.5 m in 5 s intervals. This dataset, coarsened to 10 s temporal bins, contains 7,213 contact events over 2,760 observed dyads. Participants self-reported their relationship category with respect to the wedding couple, enabling group mixing analysis. Beyond implications for the generalizability of interaction patterns, this dataset supports social event modeling for applications from contact tracing to social-space design.

## Objective

How do people, and their social groups, mix at a party? We captured a fine-grained dataset of unstructured social interaction with socially meaningful group labels to fill a gap in the study of face-to-face interaction.

Human face-to-face interaction data from academic conferences [1], classrooms [2], hospitals [3], and workplaces [4] have exhibited some common patterns such as heterogeneous contact rates [5], clustering [6], demographic homophily [7], degree inequality [8], superlinear relationships between degree and strength [5], weaker and fewer cross-group ties [9], and bursty dynamics [10]. The consistency of these patterns across varied settings is consistent with the hypothesis that they are general features of face-to-face interaction [6].

However, these data collection settings are predominantly institutionally structured [2][6][9]. Schedules, work zones, and classroom formats shape who meets whom, when, and for how long. They feature formal authority roles such as supervisors, teachers, and professors that determine the audience, timing, and tenor of conversations involving them. These institutional structures and roles may therefore obscure unstructured social group dynamics. Existing studies on group mixing during face-to-face interaction events making use of these datasets often rely on shared demographic proxies like gender [8], or assigned categories like school classes [2], [9], instead of social relationship-based groups.

We do not yet know if the network properties and dynamics from institutionally structured settings generalize to unstructured social interaction. The network science community is missing fine-grained datasets of unstructured social interaction with socially meaningful group labels, motivating our data collection.

In this Data Note, we present the first, to our knowledge, public temporal proximity network dataset of a privately hosted social event, with contextual relationship-based membership. At the outdoor cocktail hour of a wedding, 95 participants wore proximity sensor badges that detected when other badge-wearers were within approximately 1.5 meters, in five-second intervals (coarsened to 10 s resolution in data processing). Participants self-reported their relationship category with respect to the wedding couple and their families, enabling analysis of socially meaningful group mixing.

This dataset, therefore, follows the measurement approach of the fine-grained temporal network datasets collected by the SocioPatterns collaboration predominantly in institutional settings, while differing in sensing technology (Ultra-Wideband [UWB] vs. active Radio-Frequency

Identification [RFID]), temporal resolution (10 s vs. 20 s), observation window (approximately 1 hour vs. 2 days–2 weeks), and participant count (95 vs. 81-403) [1-9, 11]. The cocktail hour social context more closely resembles Ingram and Morris's 80 minute study of 92 Executive MBA participants at a mixer they convened, using earlier proximity badge technology; to our knowledge, the mixer dataset has not been publicly released [12]. Our publicly released dataset distinctively combines a social hour not convened for research, pre-existing social relationships, and relationship category labels, at 10 s resolution.

Beyond testing whether previously identified face-to-face interaction patterns generalize to unstructured social gatherings, this dataset also supports more accurate modeling of social events for applications such as contagion risk estimation [4], contact tracing policy [13], social-space design [14], or interventions to support social inclusion [15].

## Data Description

We collected high-resolution proximity data at a wedding's cocktail hour from 95 participants wearing electronic badges.

### Setting and Participants

Wedding attendees over age 18 were invited to participate. Ninety-five of approximately 220 attendees enrolled. The approximately 65 min cocktail hour followed the wedding ceremony and preceded the dinner reception. Guests mingled freely on a patio with buffet stations and a bar.

### Equipment

We deployed Fleetwood Electronics Instant-Trace (IT) electronic sensor badges that measure distance to other IT badges using Ultra-Wideband (UWB) signals with an accuracy of ±0.1 m and record contacts within a set threshold distance [16]. We set the threshold to 1.5 m to match RFID-based data collections [3], [5]. When two bodies obstruct the path of the UWB signal, such as when two people are standing back-to-back, the threshold is effectively 1 m.

A contact event is logged as soon as a badge detects another badge within the set distance on its few-second signal cycle, and concludes when the measured distance exceeds 2 m. A badge's internal software resolves the contact events into intervals on the badge's fixed 5 s grid [16].

**Participant Attributes**

Prior to the event, participants self-reported expected wedding seating affiliation via a Qualtrics single-select survey question. Response options included spouse-specific friends, relatives, and family friends, plus the spouses' mutual friends. In the released dataset, we collapse non-responses and groups under six into "other." A breakdown of participant category counts is provided in Supplementary Table S1.

**Validation**

We conducted limited bench checks of badge measurement accuracy, responsiveness, and reporting symmetry. Two limited bench checks compared badge reports with video recordings of controlled contacts to explore badge detection sensitivity and inform data processing design. These checks were not intended to estimate in-situ performance. In a latency check, a target badge recorded contact within 5 s for 1/6 contact events, within 8 s for 3/6, and within 10 s for 5/6 (one contact was not reported before the next staged contact). In a 2 min controlled detection exercise, two badges both reported contact during each of four staged contact events, in 12/30 of the badges' 5 s intervals; and 11/15 intervals with contact received a report from at least one badge. We did not encounter detection-level false positives in our bench checks. We found that continuous proximity between two badges was subject to asymmetric, discontinuous reporting (i.e. missed 5 s intervals). This motivated our decision to take the union of unilateral contact reports and to coarsen the temporal resolution, ensuring each time interval in the dataset spans at least one 5 s contiguous reporting cycle of all badges present, as described in the following subsection.

**Data Processing**

The IT badge cloud platform API provided a JSON list of contact reports with the reporting badge ID, the detected badge ID, and a time of contact. The JSON output obtained via the IT API access available to us at the time records consecutive 5 s reports with a shared timestamp (the correct contact time displayed via the cloud platform application), so we reconstruct the contact time of the $k^{th}$ record as $t_k = t_0 + 5k$ s for $k = 0,1,\ldots$, before applying the edge processing rules.

We record a temporal undirected edge between A and B at time $t$ if at least one badge recorded A-B contact with an inferred timestamp in the interval $(t - 10, t]$ on a global 10 s grid.

Multiple reports from a badge pair within a single 10 s interval are collapsed into one edge. We did not attempt to fill in missed detections or carry forward contact events across bins.

Because badge distribution continued into the cocktail hour, and badge drop-off similarly occurred gradually at the end of the cocktail hour, we define the observation window as follows. We begin the observation window at the earliest time when at least 75% of badges (72) have reported contacts and correspondingly end the observation window when fewer than 75% of badges subsequently report contacts. We chose 75% as a threshold to balance removing badge distribution and drop-off while retaining most of the event. This trimming rule results in a 58 min 40 s window of contact reports, starting 3 min 10 s into the scheduled start of the cocktail hour.

**Dataset**

We present the proximity data in the form of a discrete-time temporal edge list [17]. Each row contains a timestamp in seconds from the start of the observation window $t$, ID of node A, and ID of node B, indicating that node A and node B were in contact during the 10 s time bin ending at that timestamp. Edges are undirected. As noted in table 1, we also include a node list for the social group membership in format (node ID, group label).

**Table 1**: Overview of data files.

| **Label** | **Name of data file** | **File types** | **Data repository and identifier** |
|---|---|---|---|
| *Temporal edge list* | *temporal_edge_list.csv* | *comma-separated list, with header row* | *Zenodo, DOI: 10.5281/zenodo.20430824* |
| *Node list* | *node_list.csv* | *comma-separated list, with header row* | *Zenodo, DOI: 10.5281/zenodo.20430824* |
| *Readme* | *readme.md* | *markdown file* | *Zenodo, DOI: 10.5281/zenodo.20430824* |
| *Data processing code* | *build_temporal_network_dataset.py* | *Python file* | *Zenodo, DOI: 10.5281/zenodo.20430824* |

## Limitations

**Limitations in the measure.** Interactions with non-badge wearers (approximately 60% of guests) are not observed. Since participation is opt-in, this missing data is potentially non-random. Badge-wearers may differ systematically from non-badge wearers. Participation rates varied among relationship groups, from approximately 20% to 70% (see Supplementary Table S1); network structure and group mixing measures from the observed sub-network may not generalize to the complete network [18]. Social peers of the wedding couple enrolled at higher rates than relatives/family-friend categories, so the dataset may overrepresent within-social-circle and within-age-cohort interactions at the event. Participant self-awareness and awareness of other badge-wearers may affect behavior. Proximity does not necessarily entail interaction [19]; for example, participants are likely within the 1.5 m threshold distance when they are in food station/bar lines or standing back-to-back in a dense socializing area (as noted, the detection range for back-to-back is approximately 1 m). This dataset does not capture contextual factors such as proximity to the bar/buffet.

**Limitations in the sensing.** Venue geometry and variation in how participants wore badges due to outfit constraints affect consistency of the contact distance detection threshold. Badges have an internal sensing cycle and a 5 s reporting resolution; the 5 s grids in each badge are not aligned. Bench checks found that contact reports were subject to missed bins. Badges did not report measured distance. Results may be sensitive to data processing rules regarding badge report asymmetry and temporal aggregation.

**Limitations in the data collection.** We do not have a consistent log of when badges were picked up or dropped off. Contact isolation is not distinguished from badge inactivity. Due to a technical issue, distribution continued into the cocktail hour, and badges were returned gradually as participants exited the cocktail hour area into the reception hall. Contacts at the badge pickup/drop-off table, including with any volunteers at the table who may also be participants, are therefore not well distinguished from other interactions.

**Limitations in the group membership.** Relationships to the wedding party may not capture other, potentially more socially salient relationships and group identities. The dataset does not distinguish romantic partners. The reported groups are unbalanced in size and other attributes, such as age distribution. Privacy considerations restrict releasable participant attributes beyond group membership.

## Abbreviations

API – Application Programming Interface

JSON – JavaScript Object Notation

RFID – Radio-Frequency Identification

UWB – Ultra-Wideband

IT– Instant Trace

## Declarations

**Acknowledgments.** The authors thank Fleetwood Electronics for lending the Instant-Trace proximity badges used for data collection and for providing limited technical support related to device use. Stadlan thanks Dr. Ciro Cattuto and Dr. Brian Thompson for helpful advice on proximity sensors. Stadlan and Kahn are especially grateful to Dr. Devora Najjar, Dr. Daphne Schlesinger, Mr. Wade Miller, Dr. Gabby Lovett, Dr. Josh Fried, Mr. Jacob Goldstein, Ms. Ricki Heicklen, Mr. Alex Weinberg, and Ms. Eliana Melmed, whose assistance with electronics setup, study materials, badge distribution, on-site troubleshooting, and event logistics made this data collection possible. Stadlan and Kahn also thank the study participants for generously contributing their de-identified interaction data for research, and all wedding guests and hosts for graciously accommodating this unusual data collection in support of network science.

**Funding**. No external funding was received for this study. Fleetwood Electronics provided in-kind support through the loan of Instant-Trace proximity badges and limited technical support related to device use. The Center for Computational and Social Sciences in Health at Northwestern University (COMPASS) supported co-investigator Stadlan's time on this project and provided electronic materials for an additional component of this study.

**Competing Interests.** The authors declare no competing financial interests. Fleetwood Electronics had no role in study design, data analysis, interpretation of results, or manuscript preparation. Joshua Stadlan and Richard Kahn were the grooms at the data collection; this relationship was disclosed in the IRB-approved study protocol.

**Ethics Approval.** The Northwestern University IRB reviewed and approved this study on July 24, 2025 under ID STU00224522. All human-participant procedures were conducted in accordance with the approved protocol, Northwestern University IRB requirements, and applicable institutional guidelines and regulations governing social and behavioral human-subjects research.

**Consent to Participate.** Informed consent to participate was obtained through an IRB-approved process via a Qualtrics pre-event survey. Only investigators Stadlan and Birkett were involved in the informed consent process.

**Author Contributions.** J.Z.S. led all aspects of the study, including study design, data collection, data processing, analysis, and manuscript writing. R.B.K. contributed to study design and logistics. M.B. supervised the study, advised on study design and IRB process, and edited the manuscript. All authors reviewed and approved the final manuscript.

**Data Availability.** The data described in this Data Note can be freely and openly accessed on Zenodo under DOI:10.5281/zenodo.20430824. Please see table 1 and reference [17] for details and links to the data.